\documentstyle[12pt]{article}
\input epsf

\begin{document}

\begin{titlepage}

\begin{flushright}
UM-TH-99-08\\
October 1999
\end{flushright}
\vspace{2.0cm}

\begin{center}
\large\bf
{\LARGE\bf Momentum expansion of massive two-loop Feynman graphs 
           around a finite value}\\[2cm]
\rm
{Adrian Ghinculov\footnote{Address after October 1$^{st}$, 1999:
                           Department of Physics and Astronomy, UCLA, 
                           Los Angeles, California 90095-1547, USA}
and York-Peng Yao}\\[.5cm]

{\em Randall Laboratory of Physics, University of Michigan,}\\
      {\em Ann Arbor, Michigan 48109-1120, USA}\\[.2cm]
  
\end{center}
\normalsize

\vspace{2.0cm}

\begin{abstract}
We give an algorithm for obtaining expansions of massive two-loop
Feynman graphs in powers of the external momentum around a finite,
nonzero value of the momentum. This is based on our general two-loop 
formalism to reduce massive two-loop graphs with renormalizable 
interactions into a standard set of special functions. After the
algebraic reduction, the final results are obtained by numerical 
integration. We apply the expansion algorithm to treat the top-dependent
corrections of ${\cal O}(g^2 \alpha_s)$ to the $b$ quark self-energy
and extract its momentum expansion on-shell.
\end{abstract}

\vspace{3cm}

\end{titlepage}


\title{Momentum expansion of massive two-loop Feynman graphs 
           around a finite value}

\author{Adrian Ghinculov\footnote{Address after October 1$^{st}$, 1999:
                           Department of Physics and Astronomy, UCLA, 
                           Los Angeles, California 90095-1547, USA}
 and York-Peng Yao}

\date{{\em Randall Laboratory of Physics, University of Michigan,}\\
      {\em Ann Arbor, Michigan 48109-1120, USA}}

\maketitle

\begin{abstract}
We give an algorithm for obtaining expansions of massive two-loop
Feynman graphs in powers of the external momentum around a finite,
nonzero value of the momentum. This is based on our general two-loop 
formalism to reduce massive two-loop graphs with renormalizable 
interactions into a standard set of special functions. After the
algebraic reduction, the final results are obtained by numerical 
integration. We apply the expansion algorithm to treat the top-dependent
corrections of ${\cal O}(g^2 \alpha_s)$ to the $b$ quark self-energy
and extract its momentum expansion on-shell.
\end{abstract}


Analyses of electroweak data based on two-loop radiative corrections
have become necessary due to the precision level attained already.
The experimental precision
is expected to reach even deeper at the next generation of colliders 
in the near future. This motivated in recent years an enormous 
amount of work in handling massive two-loop Feynman graphs. In its 
general form, this is a notoriously difficult problem. The essence of
the difficulties is that these graphs usually lead to unknown 
special functions.

In a previous paper \cite{2loopgeneral}
we gave a general formalism for evaluating
two-loop Feynman diagrams for arbitrary kinematical variables, based on
analytical-numerical methods.
By using this framework, the internal mass structure and the 
external kinematics of the physical process under consideration are 
fully respected. Radiative corrections are obtained at the end 
by high accuracy numerical integration.

In this letter we would like to discuss the use of our general
massive two-loop framework
of ref. \cite{2loopgeneral} for obtaining external momentum expansions around
a finite value. There are 
situations where such momentum expansions around a finite value 
are needed in the evaluation of radiative corrections, most notably 
for mass and wave function renormalization constants. However, we note
that the algorithm described in this letter can produce the full 
momentum expansion around a finite value; it works systematically
for terms of higher order in the momentum as well.

We refer to the important process 
$Z \rightarrow b \bar b$ \cite{zdecay1}--\cite{zdecay}
to illustrate in some detail the technical aspects.
A piece of the analysis of this process involves the calculation of the 
$b$ quark wave function renormalization on-shell. This is a good example 
for the general problem of expanding an amplitude at a certain 
kinematical point, and we would like to use this short communication 
to treat this problem.  Let us begin with a broader discussion.

In a perturbative 
calculation, one is faced with a set of momentum space integrals, 
where the external momenta appear at various parts of the integrands.
The dependence on the invariants $\{s\}$ 
constructed from these external momenta
is implicit at this stage.  One may introduce Feynman 
parameters and integrate over the internal momenta, the result 
of which will be some Feynman parameter integrals, where the 
dependence on $\{s\}$ is made explicit.  In principle, one can then  
expand the integrands around some fixed values $\{s_0\}$.

At the one-loop level, this procedure can be easily followed, since 
the one-loop scalar integrals are relatively simple.  At higher loops, 
there are at least two complications to follow this route.  First of 
all, due to the proliferation of diagrams, 
one must have a systematic procedure. The Feynman 
parameter integrals due to different diagrams are most likely  
quite different, and for avoiding errors it is desirable to have a  
uniform treatment for all the graphs, the number of which 
can be quite formidable.  Secondly, the integrations 
over the Feynman parameters must be feasible, in the 
sense of fast convergence and high precision.

At the two-loop level, we developed in ref. \cite{2loopgeneral} 
a framework by which all 
two-loop processes with general renomalizable 
interactions can be written as a small set of Feynman parameter integrations 
over ten special functions $h_i$.  
It would seem that an expansion around 
$\{s_0\}$ should be an easy task.  However, this is not as straightforward 
as it appears, because the scalar integrals are made of functions of 
$\{s\}$, and a brute force expansion will result in rather unwieldy 
expressions.
What is needed is a procedure to expand the momentum integrals 
around $\{s_0\}$ before the momentum integration, in such a way that 
the final product is still a linear combination of the ten scalar 
integrals.  This will greatly facilitate systematic and rapidly 
convergent numerical work.  We shall describe such a procedure 
in the following, for two-point functions.

Our method is based on an expansion of individual propagators 
around a finite external momentum.
In the following, we use 
a simple one-loop example 
in QED to show how the procedure is carried out.  
Afterwards we extend it to two-loop and apply it to the $b$ quark
self-energy, and derive the wave function renormalization constant 
as needed for the $Z \rightarrow b \bar b$ process. 

Let us consider the self-energy of a fermion (the discussion for the
bosonic case proceeds similarly):

\begin{equation}
  \Sigma (k) = [ A(k^2) + \gamma_5 A_5(k^2) ] \gamma \cdot k
               + B(k^2) + \gamma_5 B_5(k^2) \;\;\; .
\end{equation}
We want to obtain an expansion of this expression in powers of $k^2$
around a finite value $k^2=k_0^2$. Given a set of Feynman graphs 
of a certain loop order, in the case when one is able to evaluate
them analytically, it is of course straightforward to perform a 
momentum expansion of the functions in the above expression.

However, when several finite masses and momenta are involved, 
in general such analytical solutions are not available beyond one loop.
Instead, we can obtain systematically an expansion of the integrands
by expanding the individual propagators which contain the external
momentum $k$. We introduce a deviation $\Delta_{\mu}$ of the external
momentum $k_{\mu}$ around the finite expansion point $k_{0 \mu}$:

\begin{equation}
   k_{\mu} = k_{0 \mu} + \Delta_{\mu}
\end{equation}
such that the deviation is orthogonal to $k_0$, $\Delta \cdot k_0 = 0$.
We have the freedom to make this specific choice for $\Delta$ because the 
coefficients in the expansion are independent of the direction of 
differentiation. Squaring $k_{\mu}$, one obtains $\Delta^2 = k^2-k_0^2$.

When writing down the integrand of a Feynman graph, the numerator is
already a polynomial of the external momentum $k$, while the propagators 
containing $k$ can be expanded around $k_0$:

\begin{equation}
   \frac{1}{(p+k)^2 - m^2} = 
   \frac{1}{(p+k_0)^2 - m^2} -
   \frac{\Delta^2 + 2p \cdot \Delta}{[(p+k_0)^2 - m^2]^2} +
   \frac{[\Delta^2 + 2p \cdot \Delta]^2}{[(p+k_0)^2 - m^2]^3} - \dots
\end{equation}
where $p$ is a loop integration momentum, and we used the fact that 
$\Delta \cdot k_0 = 0$.

Thus the dependence of the integrand on the deviation $\Delta$ becomes
polynomial. For each order in $\Delta$ one can further use the algorithm
described in ref. \cite{2loopgeneral} 
to calculate directly the desired coefficient.
One obtains directly the expansion of the functions 
$A(k^2)$, $A_5(k^2)$, $B(k^2)$ and $B_5(k^2)$. In particular, one is
able to extract directly the mass and wave function renormalization 
counterterms.

As a simple example, let us look at the electron mass operator at 
one-loop order in quantum electrodynamics.  The momentum space 
integral in Feynman gauge is :

\begin{equation}
 i \Sigma (k) = e^2 \int { \frac{d^np}{(2\pi)^n}
 \frac{\gamma^\mu [\gamma \cdot (p+k) + m]\gamma _\mu }{[(p+k)^2 - m^2]
 (p^2 - m_\gamma ^2)} },
\end{equation}
in which $m$ is the electron mass, and 
$m_\gamma $ is the photon mass regulator.  
The conventional 
way of introducing a Feynman parameter $x$ and integrating over $p$ gives:

\begin{equation}
\Sigma (k) = {e^2\over (4\pi)^{\frac{n}{2}}}\int dx\Gamma (-{\epsilon\over 2})
(m_k^2)^{\epsilon \over 2}[nm+(2-n)(1-x)\gamma \cdot k], 
\end{equation}
where $m_k^2 = xm^2 + (1-x)m_\gamma^2 - x(1-x)k^2$, and $n=4+\epsilon$.   
This can be expanded in $\epsilon$,
and after a $\overline {MS}$ subtraction we have:
 
\begin{eqnarray}
\Sigma (k) & = & - \frac{e^2}{16\pi^2} 
                    \int dx \left\{ m[2x+2(1+x)\ln{(m_x^2)}] \right.
\nonumber \\
 & & - (\gamma\cdot k - m) \left[ (1-x)[2+2\ln{(m_x^2)}]
     + \frac{4m^2x(1-x^2)}{m_x^2} \right] 
\nonumber \\
 & & \left. + {\cal O}((\gamma \cdot k-m)^2) \right\} \;\;\; ,
\end{eqnarray}
with $m_x^2=x^2m^2+(1-x)m_\gamma^2$.

In the second approach we just proposed, we expand the electron
propagator according to eq. 3, and write: 

\begin{equation}
  \Sigma = \Sigma _0 +\Sigma _1 +\Sigma _2 + {\cal O}(\Delta^3) \;\;\; ,
\end{equation}
where ($p=p'-xk_0$):

\begin{eqnarray}
  i\Sigma _0  & =& {e^2\over (2\pi)^4}\int dx d^n p' 
{nm+(2-n)\gamma \cdot [(1-x)k_0
+\Delta]\over (p'^2-m_x^2)^2}
\nonumber \\
i\Sigma _1 & = & -2{e^2\over (2\pi)^4}\int dx d^np'
x{2({2-n\over n})p'^2\gamma \cdot \Delta 
+\Delta ^2 [n m+(2-n)(1-x)\gamma \cdot k_0]\over (p'^2-m_x^2)^3}
\nonumber \\
i\Sigma _2 & = & 3{e^2\over (2\pi)^4}\int dx x^2 d^np' 4 p'^2 \Delta ^2
{m+ \left( \frac{2-n}{n} \right) (1-x)\gamma \cdot k_0\over (p'^2-m_x^2)^4} \;\;\; .
\end{eqnarray}

After making a $\overline {MS}$ subtraction in eqs. 8 and adding up the
results, we regain the result in eq. 6. This confirms the consistency 
of the expansion procedure. 
Note the terms $\gamma \cdot \Delta$ in eqs. 8, which 
combine with the $\gamma \cdot k_0$ terms to produce $\gamma \cdot k$ terms
in eq. 6, as they must.  This feature can be used as a check.

This expansion procedure can be extended directly at two-loop level.
The resulting coefficients of the $\Delta$ expansion can be 
further decomposed into $h_i$ functions and evaluated numerically
along the lines of ref. \cite{2loopgeneral}.

\begin{figure}
\hspace{1.cm}
    \epsfxsize = 13.cm
    \epsffile{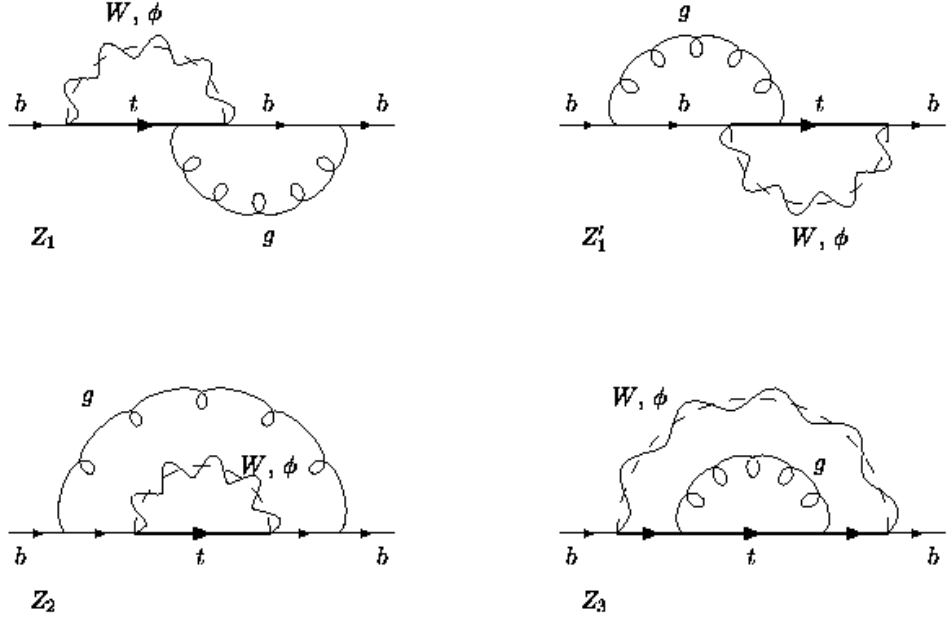}
\caption{{\em The two-loop Feynman graphs which
              contribute to the top-dependent $b$ quark self-energy 
              at ${\cal O}(g^2 \alpha_s)$.}}
\end{figure}

As an example, we give in figure 1 the two-loop Feynman graphs 
which contribute to the top-dependent $b$ quark self-energy at 
${\cal O}(g^2 \alpha_s)$. The wave function renormalization constant 
derived from this two-point function contributes to
the important process $Z \rightarrow b \bar{b}$ \cite{zdecay1}--\cite{zdecay}.

By expanding the propagators according to eq. 3, and identifying the
terms according to their power of $\Delta$, one can obtain the
momentum expansion of the functions $A(k^2)$, $A_5(k^2)$, 
$B(k^2)$ and $B_5(k^2)$ of eq. 1
on-shell, {\em i.e.} around $k_0^2=m_b^2$. The wave function 
renormalization contribution involves 
$A(m_b^2)$, $A_5(m_b^2)$, $B^{\prime}(m_b^2)$, $B^{\prime}_5(m_b^2)$, 
$A^{\prime}(m_b^2)$, and $A_5^{\prime}(m_b^2)$, where the prime denotes the
derivative with respect to the momentum squared ($\partial/\partial k^2$).

\begin{figure}
\hspace{1.cm}
    \epsfxsize = 13.cm
    \epsffile{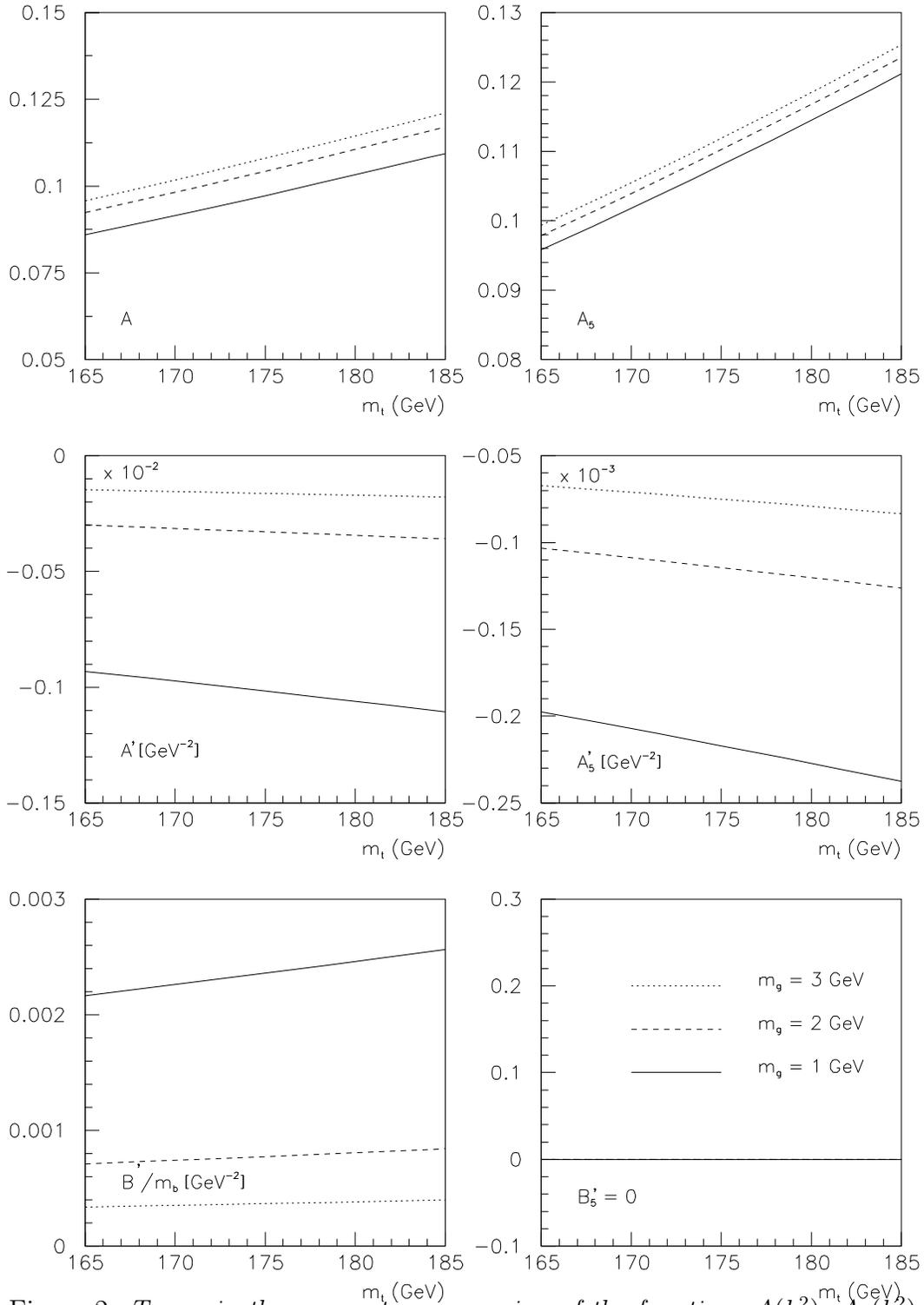}
\caption{{\em Terms in the momentum expansion of the functions
              $A(k^2)$, $A_5(k^2)$, $B(k^2)$ and $B_5(k^2)$ of eq. 1
              for the $b$ quark self-energy (figure 1). The prime denotes 
              differentiation with respect to the external momentum, 
              $\partial/\partial k^2$. The expansion is performed on-shell 
              ($k^2=m_b^2$). UV divergencies are subtracted by MS.
              We plot the results for a range of the top mass $m_t$ 
              and of the gluon infrared regulator $m_g$. 
              An overall coupling constant factor 
              of $\alpha_s (g/2\sqrt{2})^2$ 
              and a $4/3$ colour factor are understood.}}
\end{figure}

After isolating from the diagrams the
$A(m_b^2)$, $A_5(m_b^2)$, $B^{\prime}(m_b^2)$, $B^{\prime}_5(m_b^2)$, 
$A^{\prime}(m_b^2)$, and $A_5^{\prime}(m_b^2)$,
contributions, we use the general algorithm of ref. \cite{2loopgeneral} 
to evaluate them. We used $M_W=80.41$ GeV, $m_b=4.2$ GeV, and
$\sin^2\theta_W=.231$. To regularize the infrared singularities,
we introduced a mass regulator for the gluon. This being an 
${\cal O}(g^2 \alpha_s)$ correction, the use of a gluon mass regulator is
legitimate. When calculating a physical process, the gluon mass 
singularities are canceled by those from real emission diagrams; the net
effect is to replace the mass regulator by the experimental energy resolution.
The final results are given in figure 2 for a range of the top mass.
Here we used MS to treat the ultraviolet infinities.

At this point we note that in ref. \cite{fleischer} the $b$ quark self-energy
is given in the massles $b$ limit as a large top mass expansion up
to the 10$^{th}$ order. A direct comparizon of these results with 
ours is difficult because ref. \cite{fleischer} uses dimensional regularization
for handling the infrared divergencies, while we have a gluon mass regulator. 
However, we checked that in the
massles $b$ limit our result for the self-energy contains only a left-handed
component ({\em i.e.} $A(k^2) = A_5(k^2)$, $B(k^2)=B_5(k^2) \rightarrow 0$ 
in eq. 1), in
agreement with the results of ref. \cite{fleischer}. 

To conclude, we have shown how momentum expansions of massive two-loop
diagrams around a finite external momentum can be obtained systematically 
and evaluated with the general methods which we introduced previously 
in ref. \cite{2loopgeneral}. 
Such momentum expansions are needed in the computation
of radiative corrections, for instance in the evaluation of wave function
renormalization constants. The method we described in this letter has
the advantage that provides a uniform treatment of all two-loop topologies,
and is suitable for implementation in a computer algebra program for
an automatic treatment. The final result of the computer algebra program 
is a set of special functions $h_i$ which are evaluated further numerically.

\vspace{.5cm}

{\bf Acknowledgement}

This work was partially supported by the US Department of Energy (DOE).



\end{document}